\documentclass[a4paper]{jpconf}
\usepackage{graphicx}
\usepackage{epsfig}
\input symblibrary
\begin{document}
\title{First Observation of Bottom Baryon\\ 
      \( \mathbf{\Sigb} \) States at CDF}
\author{Igor V. Gorelov~\footnote[1]{talk given on behalf of the CDF
 Collaboration at the Second Meeting of the APS Topical Group on Hadronic
 Physics,~GHP~2006,~October~22~-~24,~2006,~Nashville,~Tennessee.}\\
      (For the CDF Collaboration)}
%
%
\address{Department of Physics and Astronomy,\\ 
         University of New Mexico,\\ 
         800 Yale Blvd. NE, Albuquerque, NM 87131, USA}
\ead{gorelov@fnal.gov}
%
\vspace*{-8cm}
\begin{flushright}
  {FERMILAB-CONF-07-027-E}\\
\end{flushright}
\vspace*{9cm}
%
%
\begin{abstract}
  We present the latest results on the search for bottom baryon states
  \Sigb using \( \sim1\invfb \) of CDF data.  The study is
  performed with the world's largest sample of fully reconstructed
  \Lb decays collected by CDF~II detector at 
  \( \sqrt{s}=1.96\tev \) in the hadronic trigger path. We observe 4 new
  states consistent with \Sgbstpm bottom baryons.
\end{abstract}

\section{\label{sec:Introduction} Introduction}
  High energy particle colliders provide a wealth  of
  experimental data on bottom mesons. However, only one bottom
  baryon, the \Lb, has been directly 
  observed~\cite{exp:lb-delphi,exp:lb-aleph,exp:lb-cdf1,exp:lb-cdf2}.
\par
  Heavy baryons containing one heavy quark and a light diquark became
  a nice 3-body laboratory to test QCD models. In the limit of heavy
  quark mass \( m_{Q}\to\infty \), heavy baryons' properties are governed
  by the dynamics of the light diquark in a gluon field created by the
  heavy quark acting as a static source. The heavy baryon like \Lb can
  be considered as a ``helium atom'' of QCD. In this Heavy Quark
  Symmetry (HQS) approach~\cite{th:isgur-wise}
  at the heavy quark limit a heavy quark spin does not interact with
  the gluon field, the spin decouples from the degrees of
  freedom of the light quark and the quantum numbers of the heavy quark and the light
  diquark are separately conserved by the strong
  interaction. Consequently the light diquark momentum
  \(\mathbf{j_{qq}=s_{qq}+L_{qq}}\), the heavy quark spin
  \(\mathbf{s_{Q}}\) and total momentum
  \(\mathbf{J_{Qqq}=s_{Q}+j_{qq}}\) are considered as good quantum
  numbers. Based on the HQS principles an effective
  field theory was constructed where \(\mathbf{\frac{1}{m_{Q}}}\)
  corrections can be systematically included in the perturbative
  expansions. The theory was named as Heavy Quark Effective Theory (HQET)
  (see \cite{th:hqet} and references therein).
\par
  For the bottom \(Q\equiv\,b\) baryons with a single heavy quark and two light ones 
  (see Table~\ref{tab:bbaryons}) 
  the bottom quark spin,
  \( \mathbf{s_Q={\frac{1}{2}}^{+}} \), is combined with the light diquark
  momentum \(\mathbf{j_{qq}}\) comprised of spin 
  \(\mathbf{s_{qq}=0^{+}\oplus 1^{+}}\) and its angular 
  momentum \(\mathbf{L_{qq}}\). The baryons with a diquark having \(\mathbf{s_{qq}=0^{+}}\)
  and isospin \(\mathbf{I=0}\) are called $\Lambda$- type, while the states
  with \(\mathbf{s_{qq}=1^{+}}\) and isospin \(\mathbf{I=1}\) are called 
  $\Sigma$- type. The ground state \Lb baryon has 
  \(\mathbf{I=0,\,J^{P}={\frac{1}{2}}^{+}}\).
  A doublet of ground $\Sigma$- like bottom baryons comprises 
  \Sigb with \(\mathbf{I=1,\,J^{P}={\frac{1}{2}}^{+}}\)
  and \Sigbst with \(\mathbf{I=1,\,J^{P}={\frac{3}{2}}^{+}}\). 
\par
  The combination of an orbital momentum \(\mathbf{L_{qq}=1^{-}}\) of a
  diquark with its spin of \(\mathbf{s_{qq}=0^{+}}\) adds to the
  spectroscopy a number of excited $P$- wave bottom $\Lambda$
  -states. The lowest lying orbital excitations are \Lbst with
  \(\mathbf{J^{P}={\frac{1}{2}}^{-},\,{\frac{3}{2}}^{-}}\)~\cite{Chow:1994sz,Chow:1995nw,Isgur:1999rf}.
\begin{table}[h]
\caption{\label{tab:bbaryons}Bottom baryon $\Lambda$- and $\Sigma$- states 
         and their quantum numbers. The $[q_{1}q_{2}]$ denotes 
         a pair {\it antisymmetric} in flavor and spin. 
         The $\{q_{1}q_{2}\}$ denotes a pair {\it symmetric} 
         in flavor and spin.}
\begin{center}
\begin{tabular}[h]{llll}
\bhline
{\bf State} & {\bf Quarks} & \(\mathbf{J^{P}}\) & \(\mathbf{(I,I_{3})}\) \\
\hline 
  $\Lb$      & $b[ud]$   & $(1/2)^{+}$ & $(0,0)$ \\
  $\Sigbp$   & $buu$     & $(1/2)^{+}$ & $(1,+1)$ \\
  $\Sigbz$   & $b\{ud\}$ & $(1/2)^{+}$ & $(1,0)$ \\
  $\Sigbm$   & $bdd$     & $(1/2)^{+}$ & $(1,-1)$ \\
  $\Sigbstp$ & $buu$     & $(3/2)^{+}$ & $(1,+1)$ \\
  $\Sigbstz$ & $b\{ud\}$ & $(3/2)^{+}$ & $(1,0)$ \\
  $\Sigbstm$ & $bdd$     & $(3/2)^{+}$ & $(1,-1)$ \\
  $\Lbst$    & $b[ud]$   & $(1/2)^{-}$ & $(0,0)$ \\
  $\Lbst$    & $b[ud]$   & $(3/2)^{-}$ & $(0,0)$ \\
\bhline
\end{tabular}
\end{center}
\end{table}
\par
  Theoretical expectations for ground bottom baryon states are
  summarized in Table~\ref{tab:theory}. The calculations have been
  done with non-relativistic and relativistic potential quark models,
  \(1/N_{c}\) expansion, quark models in the HQET approximation, sum
  rules, and finally with lattice QCD models~\cite{th:mass-pred}.
\par
  In a physics reality \(\mathbf{m_{Q}} \) is finite and a
  degeneration of a \(\{\Sigb,\,\Sigbst\}\) doublet is resolved by a
  hyperfine mass splitting between its states.  There is also an
  isospin mass splitting (see Table~\ref{tab:theory}) between members
  of \Sigb and \Sigbst
  isotriplets~\cite{th:lichtenberg,th:capstick,th:jrosner1998,th:jrosner}. 
  As it was pointed out~\cite{th:jrosner} the value of the isospin
  splitting within \Sigb triplet does differ from \Sigbst triplet,
  namely
  \((\Sigbstp\,-\,\Sigbstm)-(\Sigbp\,-\,\Sigbm)=0.40\pm0.07\mevcc\). 
  This number contributes to systematic uncertainty of our experimental
  results (see Section~\ref{sec:Candidates}).
\begin{table}[h]
\caption{\label{tab:theory}Mass and width predictions for \Sgbstpm.}
\begin{center}
\begin{tabular}{ll}
  \bhline
   {\bf \( \mathbf{\Sigb} \)  property } & \( \mathbf{\mevcc} \) \\
  \bhline
  \( {m(\Sigb) - m(\Lb)} \)     & \({180 - 210}\) \\
  \( {m(\Sigbst) - m(\Sigb)} \) & \({10 - 40}\) \\
  \( {m(\Sigbm) - m(\Sigbp)}\)  & \({5 - 7}\)\\
  \hline
  \( {m(\Lb)} \),{\bf ~fixed} & {\({5619.7}\)}\\
  {\bf from CDF~II} & {\( \pm1.2\pm1.2 \)}\\
  \hline
  \( {\Gamma(\Sigb),\Gamma(\Sigbst)}\) & {\( \sim8, \sim15 \)}\\
   see below & \\
  \bhline
\end{tabular}
\end{center}
\end{table}
\par
\begin{figure}[h]
\includegraphics[width=22pc]{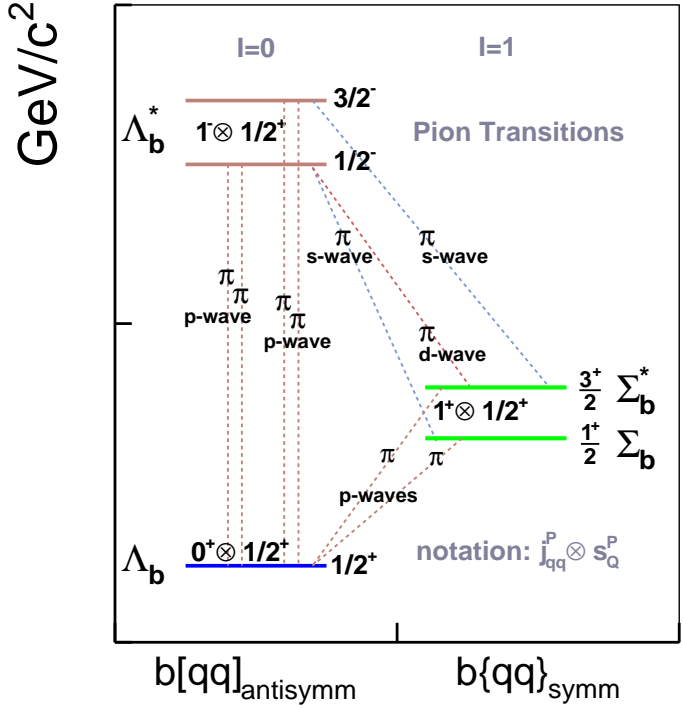}\hspace{2pc}
\begin{minipage}[b]{14pc}
\caption{
  \label{fig:exc-pions}
  Transitions of ground states \Sigb, \Sigbst into a low lying ground
  state \Lb via a single-pion emission in a $P$-wave as the diquark
  with \(j_{qq}=1^{+}\) is converted to one with
  \(j_{qq}=0^{+}\). If \Lbst states have masses sufficiently higher
  than the decay thresholds, the two-pion transitions into \Lb or a
  single-pion decays into \Sigb or \Sigbst states are possible (see
  also a discussion~\cite{Chow:1994sz,Chow:1995nw,Isgur:1999rf}).
  \Lbst single-pion transitions to \Lb are forbidden due to isospin
  conservation and also by parity (for the higher
  \(J^{P}={\frac{3}{2}}^{-}\) state).  
}
\vspace{+0.8in}
\end{minipage}
\end{figure}

  According to HQS the physics of pion transitions between heavy
  baryons is governed by the light diquark. The one or two pions are
  emitted from the light diquark while the heavy quark propagates
  unaffected by the pion emission process. Various pion transitions of
  bottom baryons into the lower ground states are summarized at the
  Figure~\ref{fig:exc-pions}. The mass predictions for the $S$- wave
  (i.e. ground state) $\Sigma$- like baryons show that there is enough
  phase space for the both \Sigb and \Sigbst to decay into \Lb via
  single-pion emission. The two excited ($P$-wave) \Lbst states might
  decay into \Lb via two-pion transitions provided a sufficient phase
  space.
\par
  It is important for our experimental expectations 
  to understand the natural width of \Sgbst baryons. As we expect 
  that \Sgbst masses lie within \((180 - 210)\mevcc\) above \Lb 
  and well above a threshold for a single-pion mode, we would expect that 
  the single-pion $P$-wave transition will dominate the total 
  width~\cite{th:peskin,th:koerner}. The authors~\cite{th:koerner}
  find  
  \[ \Gamma_{\Sigma_Q\to\Lambda_Q\pi} 
	= \frac{1}{6\pi}\frac{M_{\Lambda_Q}}{M_{\Sigma_Q}}
	\left|f_p\right|^2\left|\vec{p}_\pi\right|^{3},\] 
  where \(\vec{p}_\pi\) is the three-momentum of 
  soft \( \pi_{\Sigma_Q} \), \(f_p\equiv\,g_A/f_\pi \), \(f_\pi=92\mev \)
  and \(g_A\) is the axial vector coupling of the constituent 
  quark for the nucleon. A fit of this formula to the known PDG 
  width measurements~\cite{pdg} of charm states \Sigc and \Sigcst yields
  \(g_A=0.75\pm0.05\) which is in excellent agreement with \(g_A=0.75\)
  numerical theoretical value for the nucleon. Using the fit 
  results we have estimated \(\Gamma(\Sgbst)\), see 
  the 
  Table~\ref{tab:theory}. 
  The error of the fitted \(g_A\)
  contributes as a systematic uncertainty to our 
  experimental measurements (see Section~\ref{sec:Candidates}).
\section{\label{sec:Principle} Principle of the Analysis}
\begin{figure}[h]
\includegraphics[width=18pc]{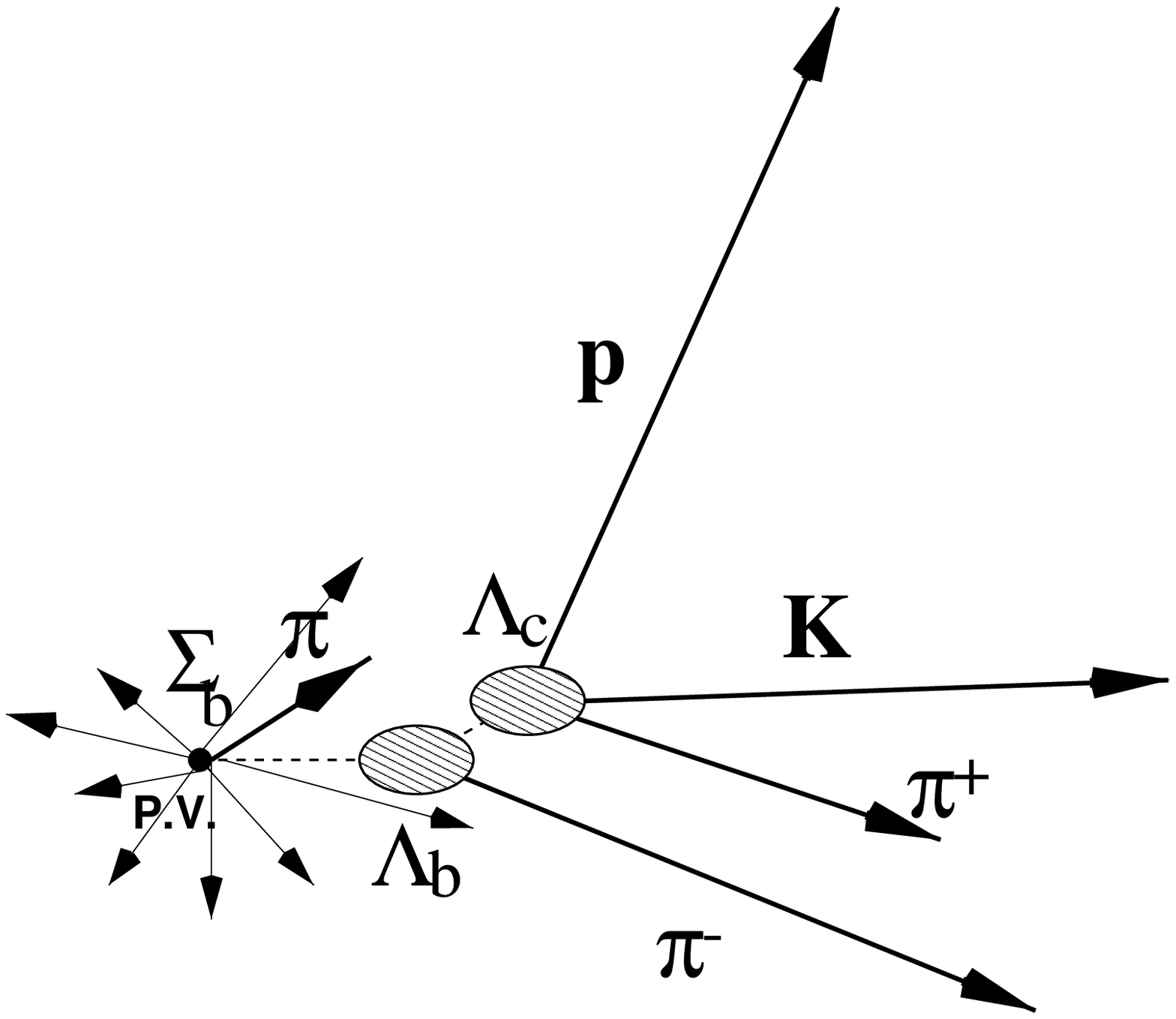}
\hspace{2pc}
\begin{minipage}[b]{18pc}
\caption{
\label{fig:topology}
  Sketch of the event topology of a \Sigb produced in the CDF
  detector~\cite{cdf:det}. The bottom baryon \Sigb is produced at the collision
  origin, i.e. primary vertex of the event.  The \Sigb decays strongly
  into \Lb with soft \(\pi_{\Sigb}^{\pm}\) emitted.  Due to a fast
  nature of the strong decay the soft pion originates from the primary
  vertex. The \Lb and its daughter \Lc decay weakly with a secondary
  decay vertex measured in SVX~II\cite{cdf:svx2}. The tracks with a
  common secondary vertex are displaced relative to a primary vertex
  and the proton from \Lc decay and \pim from \Lb decay most likely
  contribute to the CDF hadron Two Track Trigger (see
  Section~\ref{sec:Triggers} and \cite{cdf:svt}).
        }
\end{minipage}
\end{figure}

  The topology of the event with \Sgbst state produced in Tevatron collisions
  is demonstrated in Figure~\ref{fig:topology}. The \Sgbst candidates are 
  searched in the decay chain\footnote[2]{Unless otherwise stated all 
  references to the specific charge combination imply the charge conjugate 
  combination as well.}:
  \begin{itemize} 
    \item Strong decay \(\Sgbstpm\to\LambB\pi_{\Sigb}^{\pm}\) with both \(\LambB\to\Lcpim\) 
          and its daughter \(\Lc\to\pKpi\) in weak decay modes.
  \end{itemize}
  To remove a contribution due to a mass resolution of each \Lb
  candidate and to avoid absolute mass scale systematic uncertainties,
  the \Sgbstpm candidates are reconstructed in the mass difference
  {\it Q-value} spectra defined as \[Q =
  M(\LambB\pi_{\Sigb}^{\pm})-M(\Lb)-\,M_{\rm PDG}(\pipm).\] The narrow
  signatures are searched for in the $Q$-spectrum constructed
  separately for every charge state of \Sgbstpm candidates.  The
  subsample of \Sgbstm contains \( \Lb\pim \) and \(
  \overline{\Lb}\pip \) combinations from the decays of the particles
  \Sgbstm and the antiparticles \( \overline{\Sigb}^{(*)+} \),
  respectively.  The subsample of \Sgbstp contains \( \Lb\pip \) and
  \( \overline{\Lb}\pim \) combinations from the decays of the
  particles \Sgbstp and the antiparticles \( \overline{\Sigb}^{(*)-} \),
  respectively.
\par
  The \Sgbstpm signal region at $Q$- value spectrum is defined as
  \(30\mevcc\lsim Q\lsim100\mevcc\), based on the theoretical expectation
  (see Table~\ref{tab:theory}). 
  We pursued a {\it blind analysis} and
  developed the \Sgbstpm selection criteria using {\it only} the pure
  background sample in the upper and lower sideband regions of
  \(0\mevcc\lsim Q\lsim30\mevcc\) and \(100\mevcc\lsim Q\lsim500\mevcc\). 
  The signal was modeled by a {\sc PYTHIA}~\cite{th:pythia} Monte Carlo.
\section{\label{sec:Triggers} Triggers and Datasets}
  Our results are based on data collected with the \cdf2
  detector~\cite{cdf:det} and corresponding to an integrated
  luminosity of $\sim1.1\invfb$. As $\pap$ collisions at 1.96\,TeV
  have an enormous inelastic total cross-section of \(
  \sim\,60\,\mbarn \), while \b- hadron events comprise only \(
  \approx\,20\,\mub\,\,(|\eta|<1.0) \), triggers selecting \b- hadron
  events are of vital importance. Our analysis is based on a data
  sample collected by a three-level Two displaced Track Trigger.  It
  reconstructs a pair of high \( \pt>2.0\gevc \) tracks at Level~1
  with the CDF central tracker and enables secondary vertex selection
  at Level~2. This requires each of the tracks to have an impact
  parameter measured by the CDF silicon detector
  SVX~II~\cite{cdf:svx2,cdf:svt} to be larger than 120\mum.  The
  excellent impact parameter resolution of SVX~II makes this
  challenging task possible. The trigger proceeds with a full event
  reconstruction at Level~3.  The Two displaced Track Trigger is
  efficient for heavy quark hadron decay modes (see
  Figure~\ref{fig:topology} and its caption).
\section{\label{sec:Selection} Event Selection}
  The candidates of the basic state in our analysis, \Lb, have been
  reconstructed in the mode \( \Lb\to\Lcpim \) with \( \Lc\to\pKpi
  \). The Two displaced Track Trigger requirements (see
  Section~\ref{sec:Triggers}) are confirmed offline for each \Lb
  candidate. The Charm \Lc and bottom \Lb candidates have both been
  subjected to 3-dimensional vertex fits. The collection of the fitted
  \Lc candidates has been confined to a mass range of 
  \(m_{\Lc}^{PDG}\pm 16\mevcc \)~\cite{pdg}. To suppress a prompt
  background we apply a cut on the proper decay time \( \ct(\Lb)>250\mum
  \) with its significance \( \ct(\Lb)/\sigma_{\ct}>10 \). The proper
  decay time of \Lc with respect to the \Lb vertex is required to be \(
  -70\mum<\ct(\Lc\from\Lb)<200\mum \). We define the topological
  quantities as \( \ct\equiv L_{xy}\cdot m_{\Lambda_{Q}}/\pt \) and \(
  L_{xy}=\vec{D_{xy}}\cdot\vec{\pt}/\pt \) where \(
  \vec{D_{xy}},\,\vec{\pt} \) are the vectors of corresponding
  distance or momentum in a transverse plane. To reduce combinatorial
  background and contribution from partially reconstructed modes the
  impact parameter \( d_0(\Lb) \) is also restricted to be below of 
  80\mum, where \( d_{0}=|\vec{D_{xy}}\times\vec{\pt}/\pt| \). The
  kinematic cuts for \Lb and \Lc candidates, \( \pt(\Lb)>6.0\gevc \)
  and \( \pt(\Lc)>4.5\gevc \) are applied as well.
\par
\begin{figure}[h] 
\begin{minipage}{23pc}
\vspace{-3.0in}
\begin{flushleft}
\textbf{\textsf{CDF~II Preliminary, L=1.1\invfb}}\\ 
\vspace{-0.013in}
\includegraphics[width=25pc]{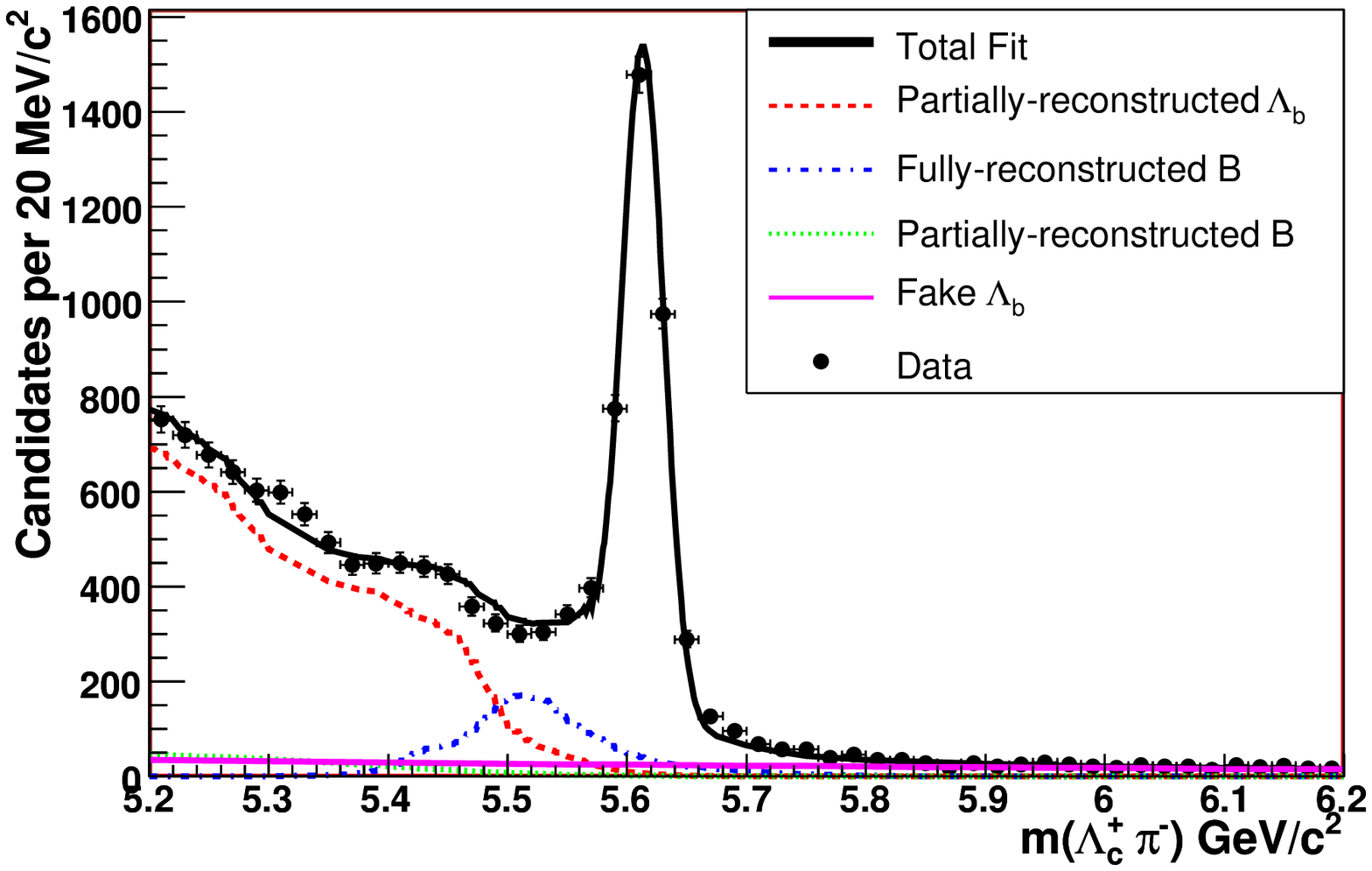}
\end{flushleft}
\end{minipage}\hspace{1.5pc} 
\begin{minipage}[b]{14pc}
\caption{
\label{fig:lb_mass}
       Fit to the invariant mass of \( \Lb\to\Lcpim \) candidates.
       Fully reconstructed \Lb decays such as \( \Lb\to\Lcpim \) and \(\Lb\to\Lc{K^{-}}\) 
       are not indicated on the figure.
       The \Lb signal region, \(5.565\gevcc < m(\Lcpim) < 5.670\gevcc\),
       consists primarily of \Lb baryons, with some contamination
       from \B mesons and combinatorial events. The left side band is 
       enriched by partially reconstructed \Lb decays like $\Lb\to\Lambda_c^{*+}\pi^-$
       and by fully reconstructed 4-prong \B- meson decays like 
       \(\Bd\to\Dp\pim,\,\Dp\to\Km\pip\pip \). The right side band consists 
       from a combinatorial background.
}
\end{minipage}
\end{figure}

  The powerful \( \Lb\to\Lcpim \) signal is shown in
  Figure~\ref{fig:lb_mass} with a binned maximum likelihood fit
  superimposed. The background from physical states contributing to
  the left side of the signal is analyzed with Monte Carlo
  simulations. The fit to the invariant \Lcpim mass distribution
  yields \( 3125\pm62\stat \) of \( \Lb\to\Lcpim \) candidates. We
  posses the world's largest \Lb sample.
\section{\label{sec:Candidates} \( \mathbf{\Sigb} \) Candidates and Signals}
  Following a method outlined in a Section~\ref{sec:Principle} the \Lb
  candidates (see a Section~\ref{sec:Selection}) from a signal region
  of \( [5.565, 5.670]\gevcc \) have been coupled with the pion tracks \(
  \pi_{\Sigb}^{\pm} \) to create \Sgbstpm candidates and the pairs  \(\Lb
  \pi_{\Sigb}^{\pm} \) have been subjected to a common vertex fit.
\par 
  The \Sgbst analysis cuts are optimized according to a {\it blind analysis}
  technique using the upper and lower sideband regions (see a Section~\ref{sec:Selection})
  of experimental $Q$- value spectrum while the signal is modeled
  by a {\sc PYTHIA}~\cite{th:pythia}.  The following kinematic and topological 
  variables are used : \(
  \pt(\Sigb) \), the soft pion track impact parameter significance \(
  \IPsig\,(\pi_{\Sigb})\), and the polar angle of the soft pion in a
  \Sigb- rest frame, \({\costhst\,(\pi_{\Sigb})\,=}\)
  \({\vec{p}_{\Sigb}\cdot\vec{p}_{\pi}^{*}/(|\vec{p}_{\Sigb}|\cdot|\vec{p}_{\pi}^{*}|)}\).
  A figure of merit is defined as $\epsilon(S_{\rm MC})/\sqrt{B}$,
  where $\epsilon(S_{\rm MC})$ is the signal efficiency measured 
  in the Monte Carlo sample and $B$ is the background
  in the signal region estimated from the upper and lower sidebands.
  The maximum of the figure of merit is reached for \(
  \pt(\Sigb)>9.5\gevc \), \( \IPsig\,(\pi_{\Sigb})< 3.0 \), and \(
  \costhst >-0.35 \). 
\begin{figure}[ht]
\includegraphics[width=18pc]{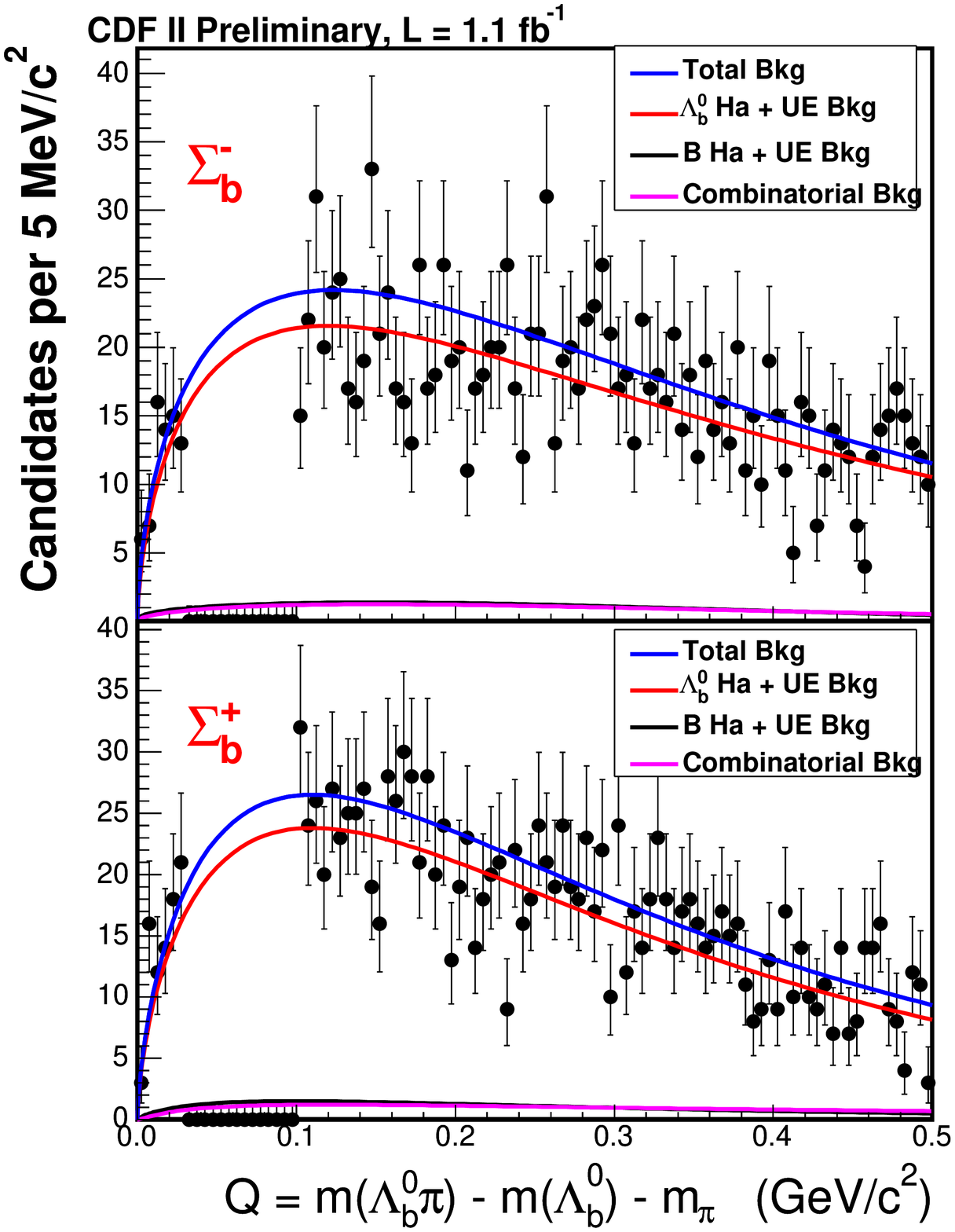}
\includegraphics[width=18pc]{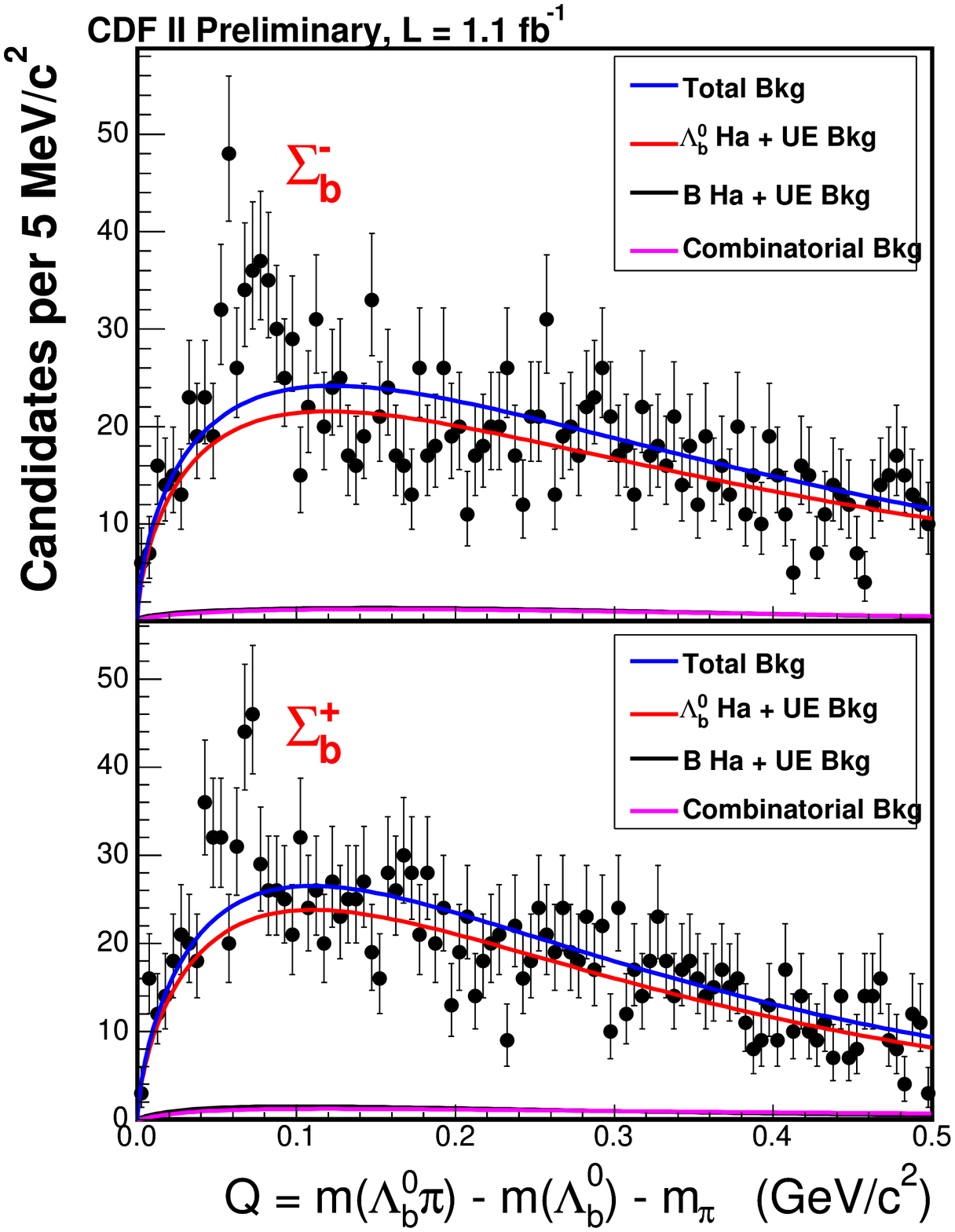}\\
\begin{center}
\vspace{-0.39in}
\textbf{\textsf{a)~~~~~~~~~~~~~~~~~~~~~~~~~~~~~~~~~~~~~~~~~~~~~~~~~~b)}} 
\end{center}
\vspace{-0.2in}
%
\caption{
\label{fig:Qblinded}
        \textbf{a)}~The three background sources described in the text
        and their sum are shown superimposed on the $Q$ distributions,
        with the signal region blinded.  The top plot shows the \(
        \LambB\pi_{\Sigb}^{-} \) distribution, which contains all
        $\Sigbm$.  The bottom plot shows the \( \LambB\pi_{\Sigb}^{+}
        \) distribution, with all $\Sigbp$. The sideband regions are
        parameterized with a power law multiplied by an exponential.
        The percentage of each background component in the \Lb signal
        region is computed from the \Lb mass fit (see
        Figure~\ref{fig:lb_mass}), and is: \(N(\Lb)\approx90.1\%\),
        \(N(B)\approx6.3\%\) and \(N({\rm
        combinatorial})\approx3.6\%\); \textbf{b)}~The excess in the
        \( \LambB\pi_{\Sigb}^{-} \) (top plot) subsample is $148$
        candidates over $268$ expected background candidates, whereas
        in the \( \LambB\pi_{\Sigb}^{+} \) (bottom plot) subsample the
        excess is $108$ over $298$ expected background candidates.  }
\end{figure}
\par
  The $Q$- value spectra with blinded signal region are shown in
  Figure~\ref{fig:Qblinded}\textrm{a} with its detailed caption.  In
  the \Sgbst search, the dominant background is from the combination
  of prompt \Lb baryons with extra tracks produced in the
  hadronization process.  The remaining backgrounds are from the
  combination of hadronization tracks with $B$ mesons reconstructed as
  \Lb baryons, and from combinatorial background events.
\par
  Upon unblinding the $Q$ signal region in both spectra we observe an
  excess of events over the background as shown in
  Figure~\ref{fig:Qblinded}\textrm{b} with the details explained in the caption.
\par
  Next we perform a simultaneous unbinned maximum likelihood fit to
  the \( \LambB\pi_{\Sigb}^{-} \) and \(\LambB\pi_{\Sigb}^{+} \)
  subsamples for a signal from each expected \Sgbstpm state plus the
  background, referred to as the ``four signal hypothesis.''  Each
  signal consists of a Breit-Wigner distribution convoluted with two
  Gaussian distributions describing the detector resolution, with a
  dominant narrow core and a small broad component for the tails.  The
  natural width of each Breit-Wigner distribution is computed from the
  central $Q$- value (see a Section~\ref{sec:Introduction} and
  \cite{th:koerner}).
  The fit shown in  Figure~5 results in the yields 
  \(N_{\Sigbp}  = 33^{+13}_{-12}\),
  \(N_{\Sigbm}  = 62^{+15}_{-14}\),
  \(N_{\Sigbstp}= 82\pm17\), and
  \(N_{\Sigbstm}= 79\pm18\) 
  candidates, with the signals located at
  \(Q_{\Sigbp}       = 48.2^{+1.9}_{-2.2}\mevcc\),
  \(Q_{\Sigbm}       = 55.9^{+1.0}_{-0.9}\mevcc\), and
  \(\Delta_{\Sigbst} = 21.5^{+2.0}_{-1.8}\mevcc\), where 
  \(\Delta_{\Sigbst}\equiv(Q_{\Sigbst}\,-\,Q_{\Sigb})\).
\begin{tabular}{c c}
\begin{minipage}[b]{18pc}
\begin{center}
  \includegraphics[width=18pc]{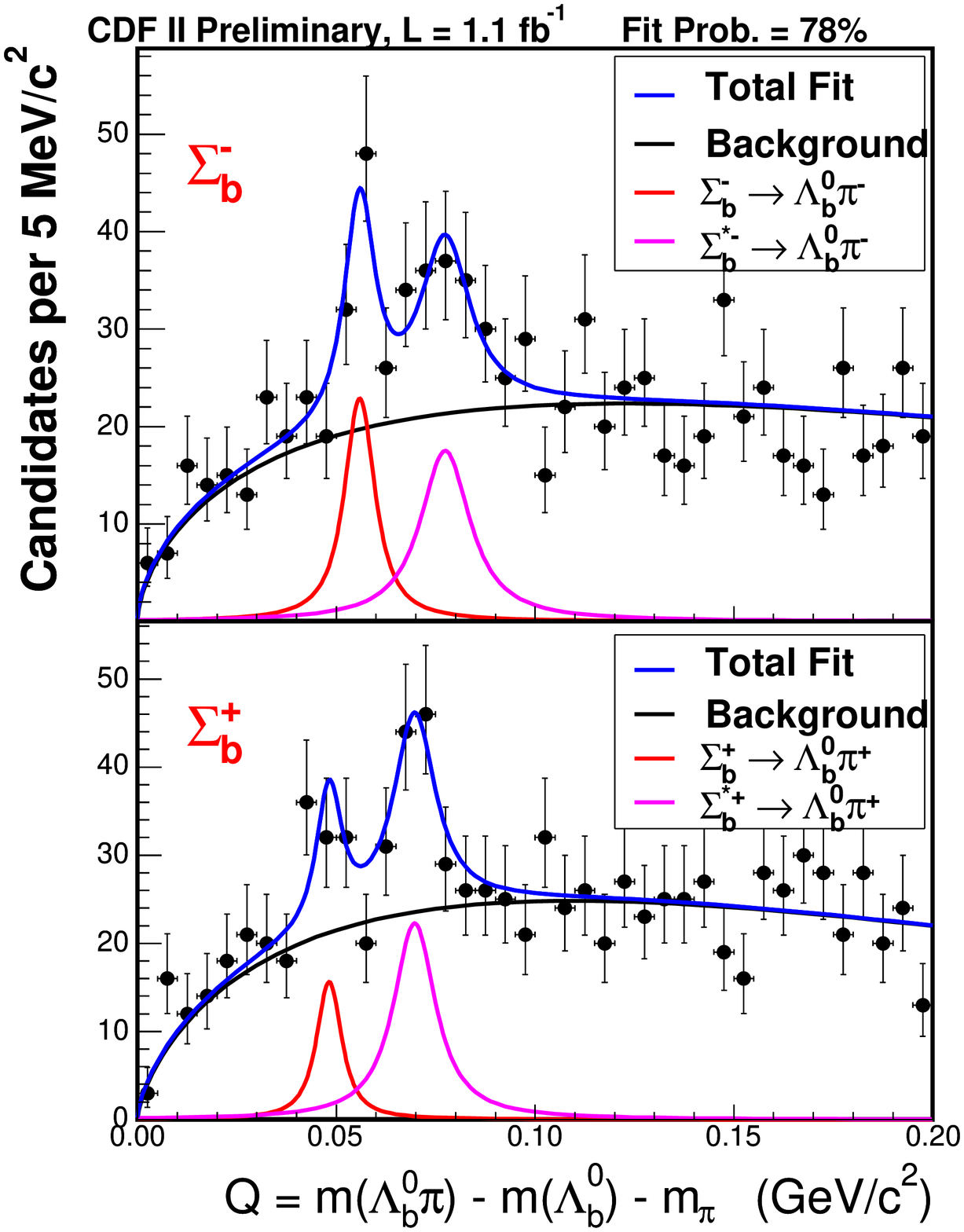}\\
  \begin{tabular}[t]{l} 
   \textbf{Figure 5.}~Simultaneous fit to the \( \LambB\pi_{\Sigb}^{-} \) \\ 
   (top) and \( \LambB\pi_{\Sigb}^{+} \) (bottom) spectra for \( {\Sigb}^{(*)\mp} \)\\
    candidates shown on a range of \\
    $Q \in$ [0, 200] $\mevcc$.
  \end{tabular}\\
\end{center}
\end{minipage}
&
\begin{minipage}[b]{18pc}
  \begin{tabular}[t]{l} 
   \textbf{Table 3.}~Likelihood ratios calculated for the \\
   alternative signal hypothesis with respect to\\
   the one of four \Sgbstpm states. The strength  of \\
   the signal hypothesis is further given  by the \\
   likelihood ratio, $LR\equiv L/L_{\rm alt}$, where $L$ is the \\ 
   likelihood of the four signal hypothesis and \\
   $L_{\rm alt}$ is the likelihood of an alternate \\
   hypothesis.
  \end{tabular}\\ 
  \begin{center}
  \begin{tabular}[b]{ll} 
  \bhline
    Hypothesis		& $LR$ \\
  \bhline
    Null	        & $2.6\times 10^{19}$	\\
    Two \Sigb    States & $1.6\times 10^{6}$	\\
     No \Sigbm   Signal	& $3.3\times 10^{4}$	\\
     No \Sigbp   Signal	& $3$			\\
     No \Sigbstm Signal	& $2.4\times 10^{4}$	\\
     No \Sigbstp Signal	& $1.8\times 10^{4}$	\\
  \bhline
  \end{tabular}\\
  \end{center}
  \begin{tabular}[t]{l} 
  The alternative hypothesis is estimated using \\
  all systematic variations of the background \\
  and signal functions. The variation with the \\
  largest value of $L_{\rm alt}$ corresponding to the least \\   
  favorable hypothesis is taken.  
  \end{tabular}
\vspace{+0.19in}
\end{minipage} 
\end{tabular}
\par
  Systematic uncertainties on the mass difference and yield
  measurements fall into three categories: mass scale, \Sgbstpm
  background model, and \Sgbstpm signal parameterization.  The mass
  scale is determined from the difference in the mean of the narrow
  resonances \(m(\Dstarp)-m(\Dz)\), \(m(\Sigczpp)-m(\Lc)\),
  \(m(\Lambda_{c}(2625)^{+})-m(\Lc)\) between data and
  PDG\cite{pdg}. The uncertainties on the background come from the
  assumption on the sample composition of the \Lb signal region, the
  normalization and functional form of the \Lb hadronization
  background.  The systematic effects related to assumptions made on
  the \Sgbstpm signal parameterization are:
  underestimation of the detector resolution in Monte Carlo,
  the accuracy of the natural width prediction from~\cite{th:koerner}, 
  and the fit constraint that 
  \((Q_{\Sigbstp}\,-\,Q_{\Sigbp})=(Q_{\Sigbstm}\,-\,Q_{\Sigbm})\)~\cite{th:jrosner}.
  All systematic uncertainties on the \(Q_{\Sgbstpm}\) mass difference
  measurements are small compared to the statistical uncertainties.
The f\mbox{}inal results~\cite{cdf:sigbweb} for the signal yields, including systematic errors, are
%
%
  \( N_{\Sigbp} = 33^{+13}_{-12}~\mbox{(stat.)} ^{+5}_{-3}~\mbox{(syst.)}\),
  \( N_{\Sigbm} = 62^{+15}_{-14}~\mbox{(stat.)} ^{+9}_{-4}~\mbox{(syst.)}\),
  \( N_{\Sigbstp} = 82^{+17}_{-17}~\mbox{(stat.)} ^{+10}_{-6}~\mbox{(syst.)}\), and
  \( N_{\Sigbstm} = 79^{+18}_{-18}~\mbox{(stat.)} ^{+16}_{-5}~\mbox{(syst.)}\).
The f\mbox{}inal results~\cite{cdf:sigbweb} for the masses are
%
%
  \( Q_{\Sigbp} = 48.2^{+1.9}_{-2.2}~\mbox{(stat.)}^{+0.1}_{-0.2}~\mbox{(syst.)}~\mevcc \),
  \( Q_{\Sigbm} = 55.9^{+1.0}_{-0.9}~\mbox{(stat.)} ^{+0.1}_{-0.1}~\mbox{(syst.)}~\mevcc \),
  and
  \( \Delta_{\Sigbst} = 21.5^{+2.0}_{-1.9}~\mbox{(stat.)} ^{+0.4}_{-0.3}~\mbox{(syst.)}~\mevcc\).
Using the CDF II measurement of 
\(m_{\Lb} = 5619.7 \pm 1.2~\mbox{(stat.)} \pm 1.2~\mbox{(syst.)}\mevcc\)~\cite{Acosta:2005mq}, 
the masses of the four states are~\cite{cdf:sigbweb}:
  \begin{flushleft}
    \( m_{\Sigbp}   = 5807.5^{+1.9}_{-2.2}~\mbox{(stat.)} \pm 1.7~\mbox{(syst.)}~\mevcc\),
    \( m_{\Sigbm}   = 5815.2^{+1.0}_{-0.9}~\mbox{(stat.)} \pm 1.7~\mbox{(syst.)}~\mevcc\),
    \( m_{\Sigbstp} = 5829.0^{+1.6}_{-1.7}~\mbox{(stat.)} \pm 1.7~\mbox{(syst.)}~\mevcc\),
    \( m_{\Sigbstm} = 5836.7^{+2.0}_{-1.8}~\mbox{(stat.)} \pm 1.7~\mbox{(syst.)}~\mevcc\),
  \end{flushleft}
where the systematic uncertainties are now dominated by the total \Lb mass uncertainty.
\par
  The significance of the signal is evaluated with two methods: using statistical 
  Monte-Carlo pseudo-experiments and comparing the likelihoods of the default
  four signal hypothesis with pessimistic alternate ones. 
  The randomly generated background samples are fit with the four signal hypothesis.
  The probability for background to produce the observed experimental 
  number of signal events or more is found to be less than $8.5\times 10^{-8}$, 
  corresponding to a significance of greater than $5.2~\sigma$. The results on 
  study of likelihood ratios are summarized in a Table~3
  and its detailed caption.
%
\section{Conclusions}
In summary, using a sample of \( \sim3100~\Lb\to\Lcpim \) candidates
reconstructed in  \(1.1\invfb\) of CDF~II data, we search for
resonant \( \Lb\pipm \) states.  We observe a significant
signal of four states whose masses and widths are consistent with those
expected for the lowest-lying
charged \( \Sgbst \) baryons: \Sigbp, \Sigbm, \Sigbstp and \Sigbstm.  
This result represents the first observation of the \( \Sgbst \) baryons.
%
\ack
  The author is grateful to his colleagues from the CDF $B$-Physics
  Working Group for useful suggestions and comments made during
  preparation of this talk. The author thanks J.~Rosner for useful
  discussions.  The author thanks S.~C.~Seidel for support of this
  work and J.~E.~Metcalfe for reading the manuscript.
%
\section*{References}
\medskip

\smallskip
%
\end{document}